# Extracting Networks of Characters and Places from Written Works with CHAPLIN

**Roberto Marazzato[1] and Amelia Carolina Sparavigna[2]**
1 *Department of Control and Computer Engineering, Politecnico di Torino, Italy*
2 *Department of Applied Science and Technology, Politecnico di Torino, Italy*

**Abstract:** We are proposing a tool able to gather information on social networks from narrative texts. Its name is CHAPLIN, CHAracters and PLaces Interaction Network, implemented in VB.NET. Characters and places of the narrative works are extracted in a list of raw words. Aided by the interface, the user selects names out of them. After this choice, the tool allows the user to enter some parameters, and, according to them, creates a network where the nodes are the characters and places, and the edges their interactions. Edges are labelled by performances. The output is a GV file, written in the DOT graph scripting language, which is rendered by means of the free open source software Graphviz.

**1. Introduction**
The analysis of social network is relevant for several theoretical and applied sciences. Among them, we find the political and the management sciences [1-3], and all the sciences based on the gathering of information. In their frameworks, some questions concerning subjects, objects, places, times and motivations are fundamental to have a plot of actual situations. As remarked in [4], these questions, usually mentioned in the news style as the Five Ws (who, what, where, when and why), are the same the scholars of Literature wish to answer after studying literary works. Any automatic text analysis, as the one we are proposing here, needs to determine what we expect to find in a novel or play, that is, the main characters about whom the work is pivoting, their interactions and also time and places and the motivations of their behaviours.
Recently, a Literary Laboratory has been created at Stanford. This laboratory aims obtaining a machine analysis of texts available in digital libraries. People in the Laboratory, founded in 2010 by Matthew Jockers and Franco Moretti, are pursuing literary researches of a digital and quantitative nature, working in a place that we could consider as a modern times version of an ancient scriptorium. One of the projects of the laboratory is the study of plots in terms of the network theory [5-7]. However, other research groups are working on projects of automatic analysis of texts and plots too, and the number of these researches is strongly increasing, due to the importance of the task. We have already mentioned Ref.4, where the authors are proposing an algorithm for inferring social networks, which is outperforming co-occurrence baselines as well as statistical classifiers. It works on an input text where personal and geographical names have been marked up previously. Social relationships are stated by dialogues and locations at which they are active, and by the sentences that attest to these claims.
Another approach was proposed in Reference 8; the used method allowed the researchers to take a systematic and wide look at a large corpus of texts. According to [8], this method, which is based on an accurate detection of face-to-face conversations, can be used like a complement of the narrower and deeper analysis performed by literary scholars. In the publication, it is also shown that as the number of characters in a novel grows, so too does the cohesion, interconnectedness and balance of their social network [8].

Besides the common scope of the previously mentioned publications, the aim of our CHAPLIN software is to provide a friendly tool, able to perform basic descriptive statistics evaluated from correlation concepts, to help the analysis of a literary text. The whole process is automated, except the phase in which human intervention is strictly needed in recognizing characters and places. As we will show, the result is a quite readable graph of character/place network.

The name CHAPLIN, CHAracters and PLaces Interaction Network, we gave to the tool is clearly inherited from that of Sir Charles Spencer Chaplin. From the tool we obtain graphs of netwroks that, as the images of the silent films acted by Charlie Chaplin, are information pictures. Like their "title cards", we add to the edges some labels to give quantitative data too.

## 2. Purpose of CHAPLIN

This software project started with the aim of creating a tool to automate the whole process of extracting a meaningful graph representing the network of characters and places from a literary text. It is clear that such automation cannot be complete: a certain amount of human intervention is needed for basic character and place recognition. Thus, another requirement for this tool is supporting literati in this task.

The required input is a set of text files stored in a folder, containing the full text of the literary work to process; their content can be viewed as a set

$$T = \{C_i; i \in (1,...,c)\} \qquad (1)$$

The first transformation to do is automatically extracting a well chosen subset of the complete set of words appearing in $T$, by means of a suitable function $X$, which at the same time splits each $C_i$ into single words and selects them under some reasonable constraint $B$, such as "take only words long more than...", "take only capitalized words", compositions of the previous and so on; let's name this set as "raw words":

$$W = \{w_j; j \in (1,...,w)\} = X(T,B) \qquad (2)$$

Clearly, $W$ contains the names of all characters and places of the work being processed, mixed with many other words; now the user begins his recognition task, in order to extract the names of characters and places, so generating the set

$$N = \{n_k; k \in (1,...,n)\} \qquad (3)$$

where each name

$$n_k = \{v_{k,p}; p \in (1,...,v), v \geq 1\} \qquad (4)$$

is a set of variants; to each name are associated, through two functions, its main variant and the property of being a character or a place:

$$\begin{aligned} name(n_k) &:= v_{k,1} \\ type(n_k) &\in \{char, place\} \end{aligned} \qquad (5)$$

The use of variants is needed as in many works the single occurrences of a certain name can vary in their form or even appear to be quite different from the main form, for linguistic and/or narrative reasons, such as case and number inflection, epithets and so on.

The next step involves finding the occurrences of each variant, grouping them under the corresponding name, so generating a frequency distribution, and the interaction matrix:

$$f(n_k); I = I(n_k, n_h) \qquad (6)$$

Each element of *I* represents how tight the character/place $n_k$ is connected to $n_h$, based on the sum of the values returned by the proximity function

$$\Pi(\Delta) : \mathbb{N} \to [0,1] \subset \mathbb{R} \qquad (7)$$

for each couple of occurrences of a variant of $n_k$ and $n_h$ in *T*. In the previous formula, $\Delta$ is the distance, in words, between the first and the second occurrence. As a proximity function,

$$\begin{aligned}&\Pi(0) = 1 \\ &\Delta_1 > \Delta_2 \Rightarrow \Pi(\Delta_1) < \Pi(\Delta_2) \\ &\exists \Delta_s : \Delta > \Delta_s \Rightarrow \Pi(\Delta) = 0\end{aligned} \qquad (8)$$

The values of *I* are then obtained by normalizing the above mentioned sums by their maximum value, so that

$$I(n_k, n_h) \in [0,1] \subset \mathbb{R} \; \forall (n_k, n_h) \qquad (9)$$

In the written work could appear a lot of secondary characters and places; in addition to that, some of them could be very weakly linked to each other. In order to avoid very complex and nearly meaningless graphs, one should select only the most relevant ones. By setting a threshold value for the elements of both *I* and *f* (let's call them $I_T$ and $f_T$, we obtain two entities which are ready to be represented through a graph:

$$N' = \{n' \in N : f(n') \geq f_T\}$$

$$I' : I'(n_k, n_h) = \begin{cases} I(n_k, n_h) \Leftarrow I(n_k, n_h) \geq I_T \wedge f(n_k) \geq f_T \wedge f(n_h) \geq f_T \\ 0 \qquad \qquad else \end{cases} \qquad (10)$$

Each element *n'* in *N'* is a node of the graph, carrying its name *name(n')*, its type *type(n')*, and its narrative strength $f(n')$. Only non null elements in *I'* result in a non oriented edge between $n_k$ and $n_h$. The constraints imposed to choose the elements from *I* into *I'* ensure that no orphan edge will be generated. Orphan nodes, representing meaningful characters or places with scarce connection to others, could occur.

## 3. Interface and output
We implemented CHAPLIN in VB.NET. Fig. 1 reports the user interface, including
- Basic commands – upper part of the window
- Commands used to extract raw words (settings, button) – just below
- Lists aiding the user in selecting raw words into names

- Commands for creating the DOT script graph (settings, button) – right part of the windows; for a better performance, two separate threshold values have been used for characters and for places

The output is a GV file, written in the DOT graph scripting language [9]. It can be rendered by means of the free open source software Graphviz [10].

**4. An example: Das Nibelungenlied**

To test the features of CHAPLIN, we decided to process the original version of the epic poem *Das Nibelungenlied*, in his original early German text (Mittelhochdeutsch). The full text, freely available online [11], consists of 39 chapters (*Âventiuren*).

Fig. 2 is the graph generated with the following values for the processing parameters:

$$\begin{aligned} \Delta_S &= 40 \\ f_T &= 0.20 \ (chars) \\ f_T &= 0.40 \ (places) \\ I_T &= 0.35 \end{aligned} \quad (11)$$

The nodes of characters and places have a different colour.
The following can be noticed:
- Many secondary characters and places don't appear in the graph.
- The protagonist of the poem is Sigfried (Sîvrit), but he doesn't obtain the best score as character; Hagen von Tronege does, as his action continues after he kills Sigfried.
- The starkest link connects Hagen to Tronege (the correspondence of this legendary town to a real place is still discussed by the experts [12]), as he is constantly named with his complete epithet.
- The river Rhine (Rìn), and the city of Worms (Wormez), the capital of the kingdom of Burgundians (at least in this poem), appear as "isolated places": this means that such geographical elements are linked to "everything and nothing", as they are always present in each chapter, but never so strongly connected to any character.

**5. Conclusion**

CHAPLIN applies basic descriptive statistics and correlation concepts to the analysis of a literary text. The whole process is automated, except the phase in which human intervention is strictly needed. The result is a well readable graph where characters and places are plotted as nodes, and the narrative connection between them corresponds to an edge. The importance of each character/place, and the strength of each plotted link are represented by means of a score varying between 0 and 1. The name we gave to this software is clearly inherited from that of Sir Charles Spencer Chaplin. The first version was released on Feb 7, 2014, exactly 100 year after the first appearance of the Tramp (Charlot). It is then an homage we paid to one of the most relevant persons of the seventh art.

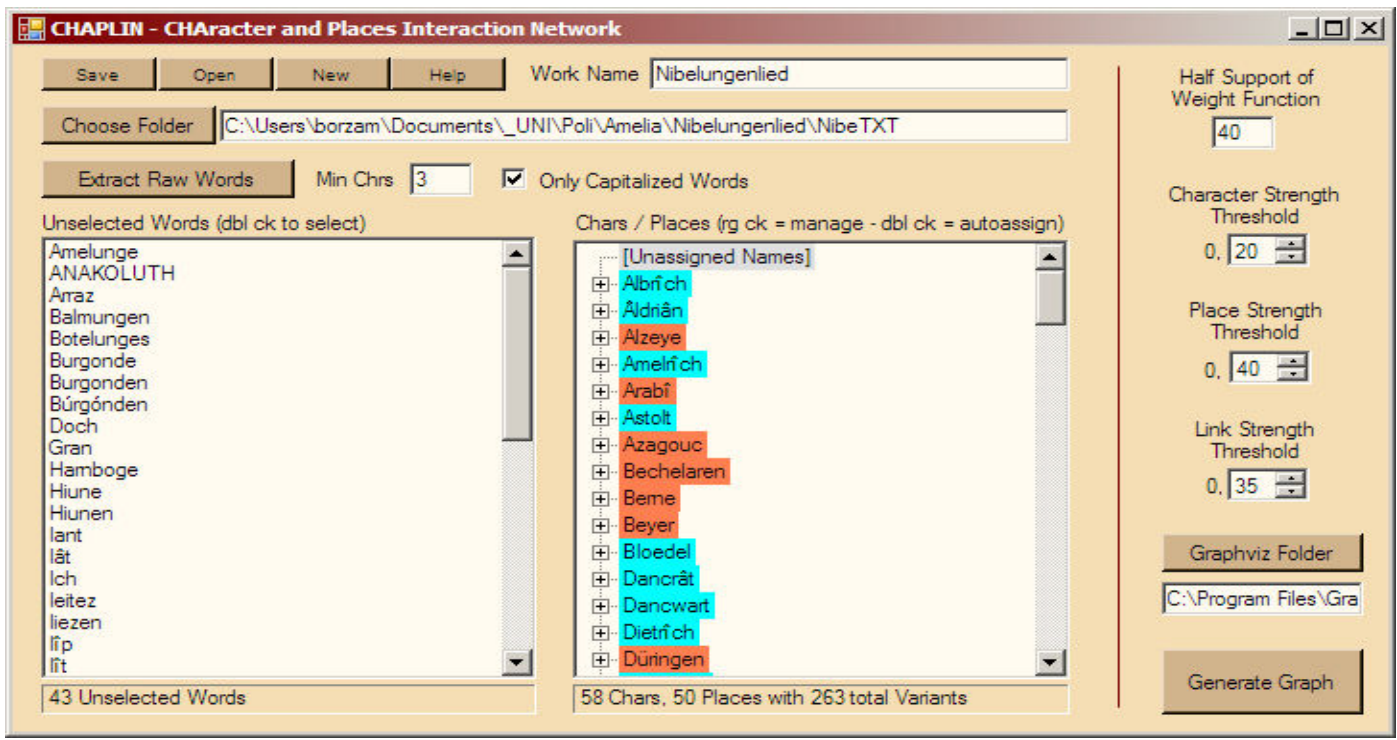

**Figure 1 – The interface of the current release of CHAPLIN.**

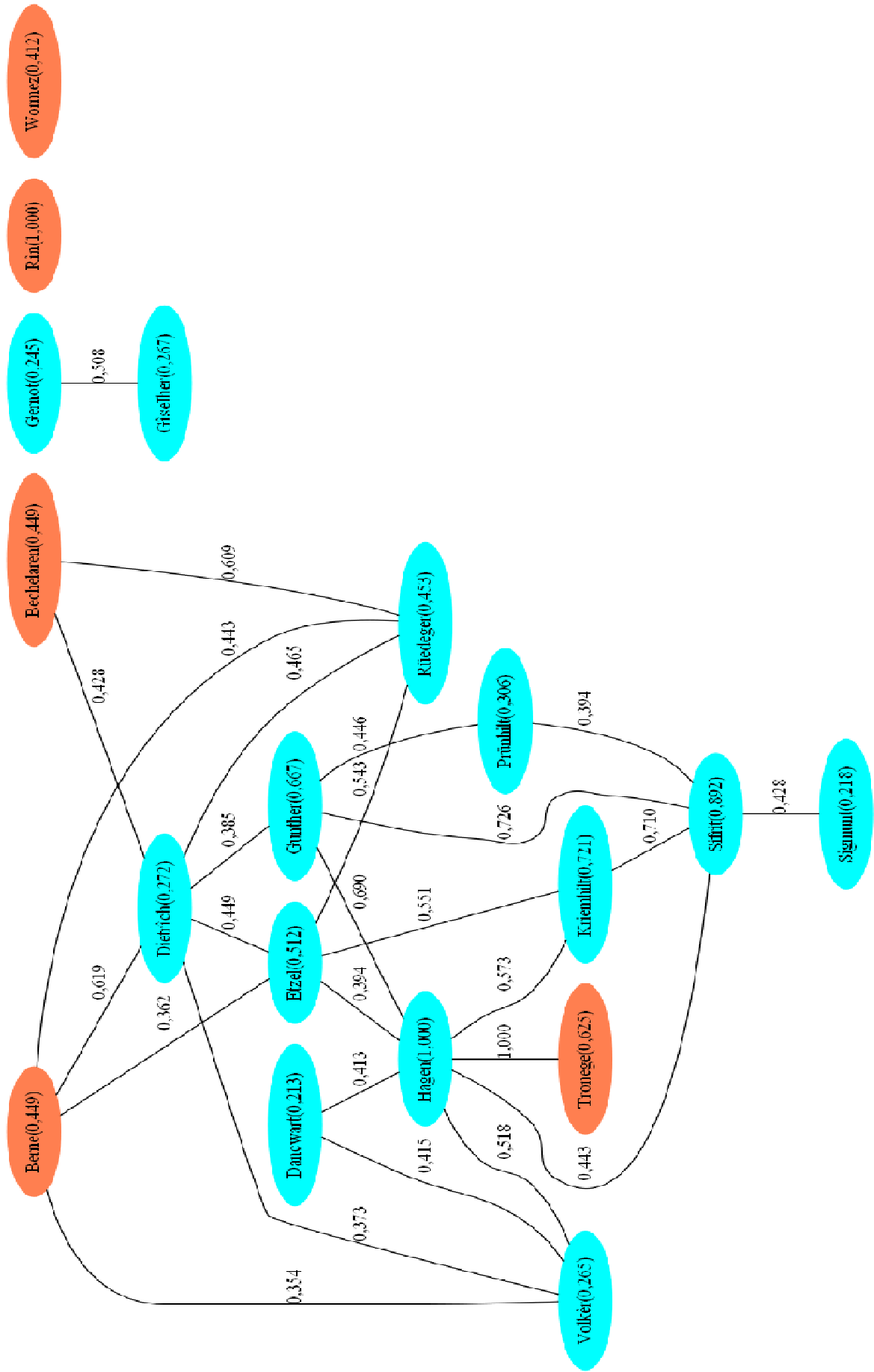

**Figure 2 –** The network obtained for *Das Nibelungenlied.*